\documentclass[apj]{emulateapj}
\usepackage{apjfonts}
\newcommand{\kms}{\,km\,s$^{-1}$}

\newcommand{\arcs}{$^{\prime\prime}$}

\newcommand{\fuse}{{\em FUSE}}

\newcommand{\iue}{{\em IUE}}

\newcommand{\hut}{{\em HUT}}

\shorttitle{FUSE Observations of H$_2$ in Starbursts}
\shortauthors{Hoopes et al.}

\begin{document}

\journalinfo{Accepted for publication in the 10 September 2004 edition of ApJ}

\title{Far Ultraviolet Observations of Molecular Hydrogen in the Diffuse Interstellar Medium of Starburst Galaxies\altaffilmark{1}}

\author{
Charles G. Hoopes\altaffilmark{2}, 
Kenneth R. Sembach\altaffilmark{3}, 
Timothy M. Heckman\altaffilmark{2}, 
Gerhardt R. Meurer\altaffilmark{2}, 
Alessandra Aloisi\altaffilmark{3,4}, 
Daniela Calzetti\altaffilmark{3}, 
Claus Leitherer\altaffilmark{3}, \& 
Crystal L. Martin\altaffilmark{5} } 

\begin{abstract}

The 905 to 1180~\AA~ spectral range of the {\it Far Ultraviolet
Spectroscopic Explorer} ({\em FUSE}) includes numerous transitions of
molecular hydrogen, making it possible to study H$_2$ in diffuse
interstellar environments directly through absorption measurements. We
have searched for H$_2$ absorption in five starburst galaxies:
NGC~1705, NGC~3310, NGC~4214, M83 (NGC~5236), and NGC~5253. We
tentatively detect weak absorption by H$_2$ in M83 and NGC~5253, and
set upper limits on the H$_2$ column density in the other
galaxies. Conservative upper limits on the mass of molecular gas
detected with \fuse\ are many orders of magnitude lower than the H$_2$
mass inferred from CO emission measurements for the four galaxies in
our sample in which CO has been detected. This indicates that almost
all of the H$_2$ is in the form of clouds with
$N$(H$_2$)$\ga10^{20}$~cm$^{-2}$ that are opaque to far-UV light and
therefore cannot be probed with far-UV absorption measurements. The
far-UV continuum visible in the
\fuse\ spectra passes between the dense clouds, which have a covering
factor $<1$. The complex observational biases related to varying
extinction across the extended UV emission in the
\fuse\ apertures prevent an unambiguous characterization of the
diffuse H$_2$ in these starbursts. However, the evidence is suggestive
that there is less H$_2$ in the diffuse interstellar medium
between the dense clouds compared to similarly reddened sight lines in the
Milky Way. This holds with the expectation that the destructive UV
radiation field is stronger in starbursts. However, previous UV
observations of these starbursts have shown that there is reddening
caused by the diffuse interstellar medium. This suggests that while
diffuse H$_2$ may be destroyed in the starburst, dust still exists.

\end{abstract}


\altaffiltext{1}
{Based on observations made with the NASA-CNES-CSA Far Ultraviolet
Spectroscopic Explorer. FUSE is operated for NASA by the Johns Hopkins
University under NASA contract NAS5-32985.}

\altaffiltext{2}
{Department of Physics and Astronomy, Johns Hopkins University, 3400
N. Charles St., Baltimore, MD 21218; choopes@pha.jhu.edu,
heckman@pha.jhu.edu, meurer@pha.jhu.edu}

\altaffiltext{3}
{Space Telescope Science Institute, 3700 San Martin Dr., Baltimore, MD
21218; sembach@stsci.edu, aloisi@stsci.edu, calzetti@stsci.edu,
leitherer@stsci.edu}

\altaffiltext{4}
{On assignment from the Space Telescope Division of ESA. Previous
address: Department of Physics and Astronomy, Johns Hopkins
University, 3400 N. Charles St., Baltimore, MD 21218}

\altaffiltext{5}
{Department of Physics, University of California at Santa Barbara,
Santa Barbara, CA 93106; cmartin@physics.ucsb.edu}

\section{Introduction}

The complex interaction between gas and stars exists in all
star-forming galaxies, but can perhaps be most easily studied in the
extreme conditions of starbursts.  The high rate of star formation in
these galaxies implies that the gravitational collapse of molecular
clouds is proceeding at a pace much greater than that seen in normal
galaxies. Once star formation has begun, the remaining gas is exposed
to the destructive UV radiation field emitted by the most massive new
stars. Illuminating the details of this relationship is a crucial step
toward understanding the evolution of galaxies through time.

Molecular hydrogen is the fuel for star formation, so it is natural to
expect large amounts of H$_2$ to exist in starburst regions. Indeed,
mm-wave observations using CO as a tracer of H$_2$ imply the presence
of $\sim3\times10^8$~M$_{\odot}$ of molecular gas in the starburst
region of M82 \citep{w01}. However, observations of some starbursts
reveal little or no CO emission, such as NGC~1705 \citep{g96} and
I~Zw~18 ({\it e.g.,} Gondhalekar et al. 1998). The lack of CO
detections is difficult to interpret for metal-poor galaxies because
the CO to H$_2$ conversion is metallicity-dependent \citep{w95}.

It is possible to study H$_2$ {\em directly} using far-UV
spectroscopy. There are numerous transitions of molecular hydrogen in
the far-UV, allowing the direct measurement of H$_2$ and eliminating
the uncertainty caused by the conversion from CO. Furthermore, far-UV
absorption studies can probe H$_2$ in column densities much lower than
can be studied through CO line emission, making it possible to study
H$_2$ in the diffuse ISM. CO studies typically detect molecular gas
with H$_2$ column densities in the range of $10^{20}$ to
$10^{23}$~cm$^{-2}$, while absorption studies have probed lines of sight in
the Milky Way and Magellanic Clouds with H$_2$ column densities from
$10^{14}$ to $10^{21}$~cm$^{-2}$ (Savage et al. 1977; Dixon, Hurwitz,
\& Bowyer 1998; Tumlinson et al. 2002). However, unlike radio
observations, far-UV measurements are profoundly affected by
extinction. In particular, absorption measurements toward extended
continuum sources with inhomogeneous extinction can be very difficult
to interpret ({\it e.g.,} Bluhm et al. 2003). Such observations must
be treated carefully in order to derive correct conclusions about the
H$_2$ content of galaxies.

\begin{deluxetable*}{lccccc}
\tabletypesize{\small}
\tablewidth{0pc}
\tablecaption{Properties of the Starburst Galaxies in the Sample} 
\tablehead{
\colhead{Galaxy} & \colhead{Type\tablenotemark{a}} & \colhead{Distance\tablenotemark{b}} & \colhead{Radial Velocity\tablenotemark{c}} & \colhead{12+log(O/H)\tablenotemark{d}} & \colhead{E(B-V)\tablenotemark{e}}\\
\colhead{}       & \colhead{}                      & \colhead{(Mpc)}                     & \colhead{(\kms)}                           &    \colhead{}                          & \colhead{}   }
\startdata
NGC~1705 & Irr AM & 5.1 & 569 & 8.0 & 0.0\\ 
NGC~3310 & SAB(r)bc & 14.5 & 1018 & 9.0 & 0.3\\ 
NGC~4214 & IAB(s)m & 2.9 & 298 & 8.2 & 0.2\\
M83(NGC~5236) & SBc & 4.5 & 503 & 9.3 & 0.3\\ 
NGC~5253 & Im Am & 3.3 & 416 & 8.2 & 0.2\\
\enddata
\tablenotetext{a}
{Galaxy Hubble types were taken from Kinney et al. 1993}
\tablenotetext{b}
{References for distances are: Tosi et al. 2001 for NGC~1705,
Ma{\'{\i}}z-Apell{\' a}niz, Cieza, \& MacKenty 2002 for NGC~4214, Thim
et al. 2003 for M83, and Gibson et al. 2000 for NGC~5253. The distance
to NGC~3310 was derived assuming $H_0=70$~km~s$^{-1}$~Mpc$^{-1}$}
\tablenotetext{c}
{Radial velocities were taken from de Vaucouleurs et al. 1991 except
NGC~1705, which was taken from Heckman et al. 2001a.}
\tablenotetext{d}
{Abundances in the inner H~II regions. References for 12+log(O/H)
values are taken from the compilation in Heckman et al. (1998).  }
\tablenotetext{e}
{The estimated intrinsic reddening (excluding the foreground Milky
Way). These are based on the measurements of the far-UV spectral
energy distribution (Meurer, Heckman, \& Calzetti 1999; Leitherer et
al.  2002), and the Calzetti (2001) effective starburst attenuation
law.}
\end{deluxetable*} 

Far-UV absorption studies have shown that H$_2$ is common in the
diffuse ISM when the shielding along the sight line is sufficient to
prevent dissociation by the interstellar UV radiation field. In the
Milky Way, the transition occurs at $E(B-V)\sim0.08$, and sight lines
with at least this much reddening almost always contain H$_2$
\citep{s77}. Investigation of the relationship between $N$(H$_2$) and
$E(B-V)$ in environments that differ from the Milky Way, such as the
intense radiation environment of starbursts, can shed light on the
formation and destruction mechanisms for H$_2$ and dust. Previous
\fuse\ investigations of H$_2$ absorption in metal-poor starbursts have
set low upper limits on the amount of H$_2$ in the diffuse ISM
(Vidal-Madjar et al. 2000; Thuan, Lecavelier des Etangs, \& Izotov
2002; Aloisi et al. 2003), and since these galaxies have not been
detected in CO it is possible that they contain very little H$_2$ in
denser environments as well. It is not clear whether the low metal
content or the starburst radiation environment is responsible for the
lack of H$_2$.  To fully understand the behavior of H$_2$ in the
diffuse ISM of starbursts, the sample must be extended to galaxies
with higher metal content and to those in which CO emission has been
detected.

We have used the {\em Far Ultraviolet Spectroscopic Explorer} ({\em
FUSE}; Moos et al. 2000) to search for H$_2$ absorption in five
starburst galaxies: NGC~1705, NGC~3310, NGC~4214, M83 (NGC~5236), and
NGC~5253. The properties of these galaxies are given in Table 1. Four
of these galaxies have been detected in CO emission, while NGC~1705
has not. The \fuse\ observations use the OB stars in the starburst as
the UV continuum source, and probe absorption by gas in front of the
starburst. The \fuse\ spectra show surprisingly little absorption from
H$_2$ in the diffuse ISM, which may be a result of observational
selection effects, but may also indicate the H$_2$ is destroyed in
starbursts.

\section{Observations and Data Reduction}

\begin{deluxetable*}{lcccc}
\tabletypesize{\small}
\tablewidth{0pc}
\tablecaption{Log of Observations} 
\tablehead{
\colhead{Galaxy} & \colhead{Data Set ID} & \colhead{Observation Date} & \colhead{$T_{exp}$} & \colhead{Aperture}  \\
\colhead{}       & \colhead{}         &    \colhead{}   &    \colhead{(ks)} &    \colhead{}        }
\startdata
NGC~1705         & A0460102,A0460103 & 2000 Feb 4-5 & 21.3 & $30^{\prime\prime}\times30^{\prime\prime}$  \\ 
NGC~3310         & A0460201          & 2000 May 5    & 27.1 & $30^{\prime\prime}\times30^{\prime\prime}$  \\ 
NGC~4214         & A0460303          & 2000 May 12    & 20.7 & $30^{\prime\prime}\times30^{\prime\prime}$  \\ 
M83 (NGC~5236)   & A0460505          & 2000 Jul 6    & 26.5 & $30^{\prime\prime}\times30^{\prime\prime}$ \\ 
NGC~5253         & A0460404          & 2000 Aug 7    & 27.4 & $4^{\prime\prime}\times20^{\prime\prime}$  \\ 
\enddata
\end{deluxetable*} 

A log of the \fuse\ observations is given in Table 2, and more
information can be found in \cite{hsmlcm01}. The \fuse\ mission and
instrument are described by \cite{m00} and \cite{s00}. The spectra
were obtained through the LWRS (30\arcs $\times$ 30\arcs) apertures
except for NGC~5253, which was observed through the MDRS (4\arcs
$\times$ 20\arcs) apertures.  Figures $1-5$ show the location of the
\fuse\ apertures for each galaxy.

The raw spectra were processed through the \fuse\ calibration pipeline
(CALFUSE v2.1.6).  The pipeline
screens data for passage through the South Atlantic Anomaly and low
Earth limb angle pointings and corrects for thermal drift of the
gratings, thermally-induced changes in the detector read-out
circuitry, and Doppler shifts due to the orbital motion of the
satellite. Finally, the pipeline subtracts a constant detector
background and applies wavelength and flux calibration. After the
pipeline reduction the individual spectra were co-added to produce the
final calibrated spectrum.

\fuse\ consists of 4 co-aligned optical channels, two optimized for
longer wavelength (LiF1 and LiF2; 1000 -- 1187~\AA) and two optimized
for shorter wavelengths (SiC1 and SiC2; 905 -- 1100~\AA). The data
from the four channels were analyzed separately, because there are
slight differences in the spectral resolution between channels which
can cause problems if the channels are co-added. Treating the channels
separately also provides a safeguard against detector defects. The
velocity zero-point of each individual channel was found using strong
Galactic absorption lines.

\section{Notes on the Individual Galaxies}

Several of the starbursts are essentially point sources in the far-UV,
and the interpretation of their spectra is fairly simple. Others,
however, are extended far-UV sources with varying extinction. In these
sources the morphology and extinction must be taken into account in
order to understand the information in the \fuse\ spectra. In this
section we describe the far-UV morphology of each galaxy and how it
may affect the results. We also review previous results of CO emission
measurements, in order to contrast them with the UV absorption
measurements presented in the next section.

{\it NGC~1705:} The strongest far-UV source in NGC~1705 is the central
star cluster NGC~1705-1, although there is a significant contribution
from other far-UV sources \citep{m95}. The galaxy is small enough that
the entire UV-emitting region fits within the \fuse\ LWRS
aperture. Figure 1 shows an archival {\it Hubble Space Telescope
(HST)} WFPC2 U-band (F380W) image and an archival FOC image
(F220W). The bright point-like nature of NGC~1705-1 is illustrated by
the point spread function apparent in the pre-costar FOC
image. NGC~1705 is metal-poor (Meurer et al. 1992; Storchi-Bergmann,
Calzetti, \& Kinney 1994; Heckman et al. 1998, 2001a) and has not been
detected in CO \citep{g96}.

\begin{figure}
\epsscale{1.0}
\plotone{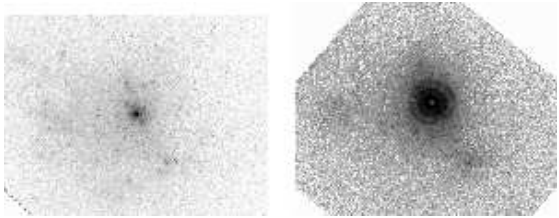}
\caption{
Ultraviolet images of NGC~1705. {\it Left:} An archival WFPC2 U-band
(F380W) image. This image is approximately
$21^{\prime\prime}\times17^{\prime\prime}$, so everything in the image
falls within the LWRS aperture. {\it Right:} An archival FOC near-UV
(F220W) image. The galaxy is essentially a point source in the UV, so
the pre-costar point spread function of {\em HST} is apparent. North
is up and East is left in both images. }
\end{figure}

{\it NGC~3310:} Figure 2 shows an {\em Ultraviolet Imaging Telescope
(UIT)} far-UV (B1 filter: 1520~\AA) image of NGC~3310, and a FOC
near-UV image (F220W) of the central region. The image shows that most
of the star-forming regions in this galaxy fit within the LWRS
aperture. A UV-bright core is surrounded by a star-forming
ring. \cite{con00} showed that the nucleus is actually quite red
compared to the ring, indicating that it is either very dusty or is
dominated by older stars. Using the UIT image shown in Figure 2,
\cite{smith96} concluded that the nucleus is heavily extinguished in
the far-UV, and that the starburst ring is the strongest far-UV
source.  The far-UV light in the \fuse\ spectrum is a composite of all
of these star-forming regions, but it is weighted toward the bluer
regions in the ring and the UV-bright knot in the southwest (known as
the ``jumbo'' \ion{H}{2} region, Balick \& Heckman 1981). There is
molecular gas traced by CO in this knot, but the highest concentration
of CO is in the northwest part of the ring
\citep{k93}. \cite{mulder95} found $\sim10^8$~M$_{\odot}$ of H$_2$,
most of which is located within the region probed by the \fuse\
pointing.

\begin{figure}
\epsscale{1.0}
\plotone{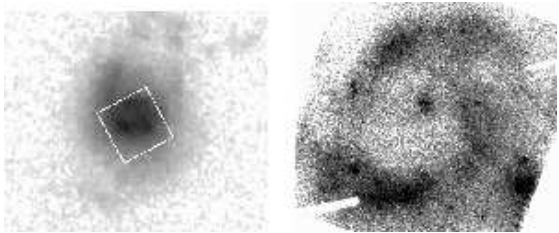}
\caption{
Ultraviolet images of NGC~3310. {\it Left:} A UIT 1520~\AA\ (B1) image
(Smith et al. 1996), showing the location of the \fuse\
$30^{\prime\prime}\times30^{\prime\prime}$ LWRS aperture. {\it Right:}
An archival FOC near-UV (F220W) image of the starburst region. The
image is approximately $28^{\prime\prime}\times24^{\prime\prime}$. The
LWRS aperture is centered south of the nucleus, at approximately the
position of the star-forming ring. North is up and East is left in
both images.}
\end{figure}

{\it NGC~4214:} The left panel of Figure 3 is a {\em UIT} far-UV
image of NGC~4214, with the \fuse\ LWRS aperture located on the
central starburst (NGC~4214-I). The right panel shows an archival
WFPC2 far-UV (F170W) image of the central region. The strongest far-UV emission
comes from NGC~4214-I, so the \fuse\ spectrum
is essentially a spectrum of this cluster. This source lies in an
extended component of diffuse CO emission, containing
$\sim8\times10^{5}$~M$_{\odot}$ of H$_2$
\citep{walter01}, and is in a ``hole'' in the H$\alpha$
emission \citep{l96}.

\begin{figure}
\epsscale{1.0}
\plotone{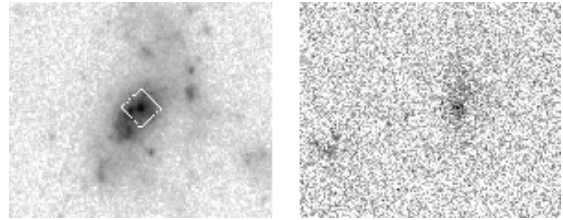}
\caption{
Ultraviolet images of NGC~4214. {\it Left:} A UIT 1520~\AA\ (B1)
image, showing the location of the \fuse\
$30^{\prime\prime}\times30^{\prime\prime}$ LWRS aperture. {\it Right:}
A WFPC2 far-UV (F170W) image of the starburst region. The image is
approximately $21^{\prime\prime}\times17^{\prime\prime}$, so
everything in the image falls within the LWRS aperture. North is up
and East is left in both images.}
\end{figure}

{\it M83 (NGC~5236):} The left panel of Figure 4 is a UIT far-UV
(1520~\AA) image of M83, with the \fuse\ LWRS aperture located on the
central starburst. The right panel shows an archival WFPC2 U-band
(F300W) image of the central region. \cite{h01} used this image to
show that the central region contains at least 39 star clusters less
than 10~Myr old. These clusters should have strong far-UV
emission, so the \fuse\ spectrum is a combination of many individual
sight lines.  \cite{israel01} found that the central region (roughly
coincident with the LWRS aperture) contains $\sim10^7$~M$_{\odot}$ of
H$_2$ from observations of CO emission.

\begin{figure}
\epsscale{1.00}
\plotone{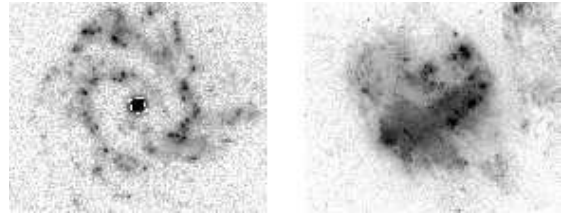}
\caption{
Ultraviolet images of M83. {\it Left:} A UIT 1520~\AA\ (B1) image,
showing the location of the \fuse\
$30^{\prime\prime}\times30^{\prime\prime}$ LWRS aperture. {\it Right:}
A WFPC2 U-band (F300W) image of the starburst region of M83. The image
is approximately $21^{\prime\prime}\times17^{\prime\prime}$, so
everything in the image falls within the LWRS aperture. North is up
and East is left in both images.}
\end{figure}

\begin{figure}
\epsscale{1.0}
\plotone{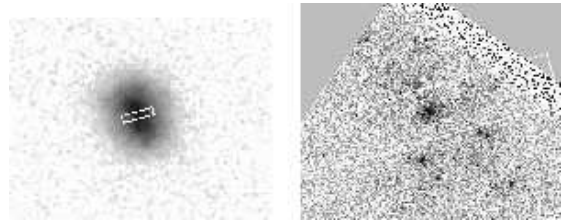}
\caption{
Ultraviolet images of NGC~5253. {\it Left:} A UIT 1520~\AA\ (B1)
image, showing the location of the \fuse\
$4^{\prime\prime}\times20^{\prime\prime}$ MDRS aperture. {\it Right:}
A WFPC2 far-UV (F170W) image, also showing the location of the {\em
FUSE} MDRS aperture. North is up and East is left in both images.}
\end{figure}

{\it NGC~5253:} This is the only galaxy for which the MDRS aperture
was used. Figure 5 shows the placement of the aperture on a UIT far-UV
image and an archival WFPC2 far-UV image. The region observed includes
the brightest cluster in the far-UV \citep{m95}, but no other regions
of significant star formation. \cite{mtb02} found that this region
contains $\sim10^5$~M$_{\odot}$ of H$_2$ from observations of CO
emission. Soft X-ray emission is associated with the cluster,
suggesting the presence of a superbubble \citep{ss99}.

\section{Column Density Measurements}

\subsection{Molecular Hydrogen}

\begin{figure*}
\epsscale{1.05}
\plotone{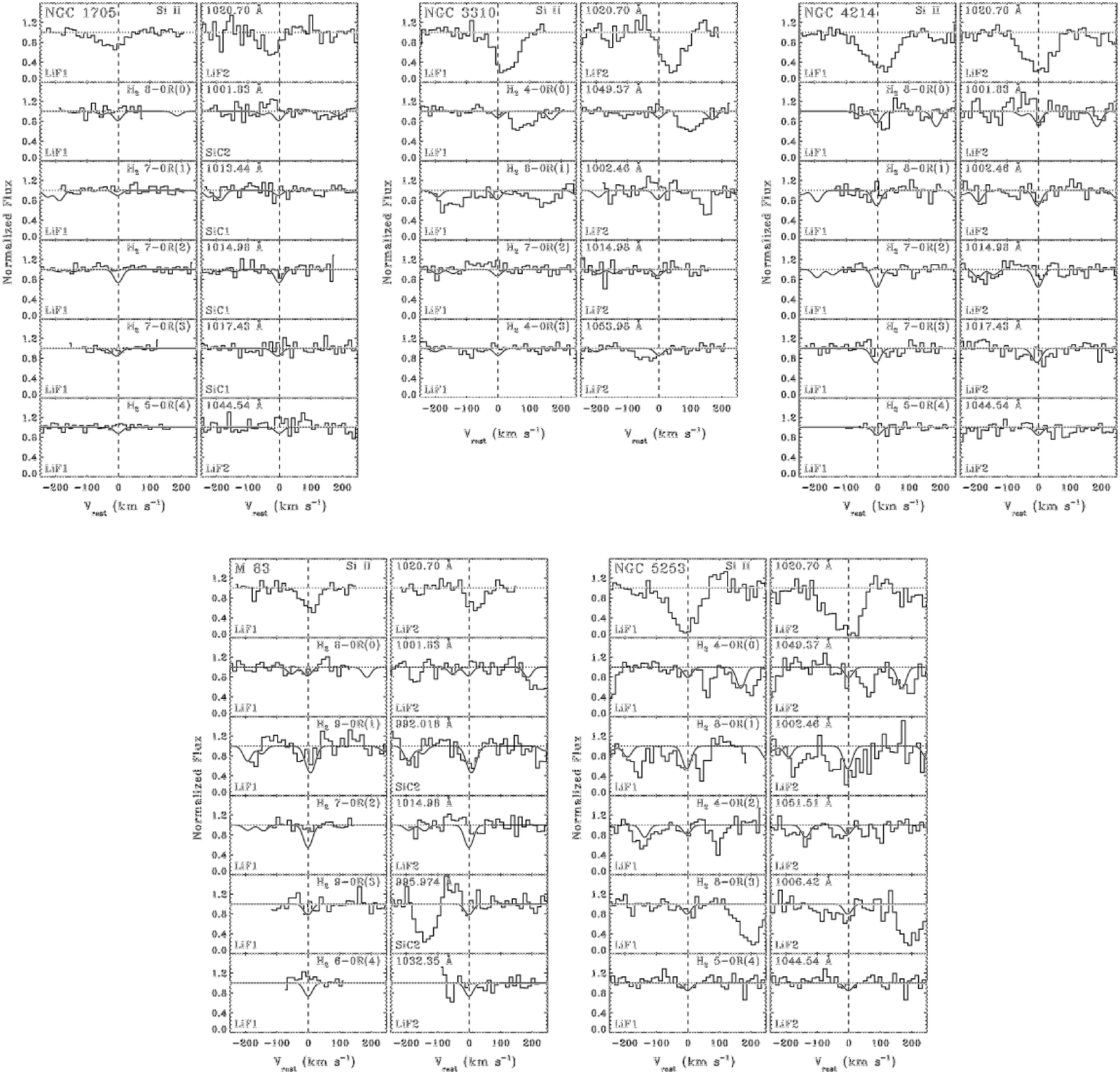}
\caption{
Interstellar absorption line profiles from the $FUSE$ spectra of the
five starburst galaxies. The histogram is the observed spectrum, and
the smooth line is a model H$_2$ spectrum based on the column
densities listed in Table 3. For each galaxy we show the
Si~II~1020.699~\AA\ profile and the expected locations of H$_2$ lines
from rotational levels $J=0-4$ (except NGC~3310, for which no suitable
$J=4$ line was found). Two channels are shown for each line, although
the measurements were made on the channel with the highest S/N
(LiF1). The velocity scale has been set so that absorption intrinsic
to the galaxy falls at 0 \kms, using the velocities listed in Table
1. }
\end{figure*}

Absorption profiles for each galaxy are shown in Figure 6. Each panel
in the figure shows the \ion{Si}{2} 1020.699~\AA\ absorption profile
(which traces the neutral ISM) as well as one H$_2$ absorption profile
for each rotation level from $J=0-4$ (except for NGC~3310, for which
no suitable $J=4$ line was found). The lines shown were chosen because
they are the strongest lines that are not blended with any other
absorption, either intrinsic or Galactic. Absorption from H$_2$ is
only visible in the $J=1$ and $J=2$ levels in NGC~5253 and in the
$J=1$ level of M83. The NGC~5253 detections in the LiF1 spectrum are
both $\sim3\sigma$ or better, but the absorption is not obvious in the
less-sensitive LiF2 channel, so the detections in LiF1 are
tentative. There is also a possible detection in the $J=3$ level for
NGC~5253, but this would be at $<2\sigma$ significance and is not
confirmed in the LiF2 spectrum so we do not count it as a
detection. The M83 $J=1$ detection is only $1.7\sigma$ in LiF1, but it
is confirmed in the LiF2 spectrum, so we count it as a detection. For
all the other galaxies we only measured upper limits.

For the galaxies in which we detected H$_2$ absorption, we measured
the equivalent widths of the strongest lines and transformed these to
H$_2$ column densities, assuming the absorption is optically thin. For
the other galaxies we measured the noise in the spectrum at the
expected location of H$_2$ absorption, and used this to compute
$3\sigma$ upper limits on the amount of H$_2$ present. All
measurements were made on the LiF1 spectrum because it is the most
sensitive. The measured equivalent widths of the strongest lines and
the derived H$_2$ column densities or upper limits are given in Table
3. A model H$_2$ spectrum was created using the column densities for
each $J$ level listed in Table 3 and assuming a Doppler broadening
parameter of $b=15$~\kms. The model is shown as the smooth black line
in Figure 6.

\begin{deluxetable}{lccc}
\tablecolumns{4}
\tabletypesize{\scriptsize}
\tablewidth{0pc}
\tablecaption{Diffuse H$_2$ Content of the Sample Galaxies\tablenotemark{a}} 
\tablehead{
\colhead{Line ID} & \colhead{Wavelength} & \colhead{Equivalent Width} & \colhead{N(H$_2$)}\\
\colhead{}       & \colhead{(\AA)} & \colhead{(m\AA)} & \colhead{($10^{14}$ cm$^{-2}$)}}
\startdata
NGC 1705 & &  & \\
8-0 R(0) & 1001.83 & $<23.0$ & $<0.97$ \\
7-0 R(1) & 1013.44 & $<12.2$ & $<0.66$ \\
7-0 R(2) & 1014.98 & $<14.6$ & $<0.85$ \\
7-0 R(3) & 1017.43 & $<11.9$ & $<0.72$ \\
5-0 R(4) & 1044.54 & $<10.2$ & $<0.70$ \\
total\tablenotemark{b} &  & & $<3.90$ \\
NGC 3310 & &  & \\
4-0 R(0) & 1049.37 & $<15.6$ & $<0.69$ \\
8-0 R(1) & 1002.46 & $<22.5$ & $<1.38$ \\
7-0 R(2) & 1014.98 & $<15.9$ & $<0.92$ \\
4-0 R(3) & 1053.98 & $<17.1$ & $<1.34$ \\
total\tablenotemark{b} & & & $<4.33$ \\
NGC 4214 & &  & \\
8-0 R(0) & 1001.83 & $<29.4$ & $<1.24$ \\
8-0 R(1) & 1002.46 & $<41.4$ & $<2.54$ \\
7-0 R(2) & 1014.98 & $<22.8$ & $<1.32$ \\
7-0 R(3) & 1017.43 & $<29.1$ & $<1.76$ \\
5-0 R(4) & 1044.54 & $<19.2$ & $<1.33$ \\
total\tablenotemark{b} & & & $<8.19$ \\
M83 & &  & \\
8-0 R(0) & 1001.83 & $<22.11$ & $<0.93$ \\
9-0 R(1) & 992.018 & $25.6\pm15.0$  & $1.62\pm0.95$ \\
7-0 R(2) & 1014.98 & $<22.9$ & $<1.30$ \\
9-0 R(3) & 995.974 & $<29.1$ & $<1.90$ \\
6-0 R(4) & 1032.35 & $<35.7$ & $<2.20$ \\
total\tablenotemark{b} &  & & $<8.90$ \\
NGC 5253 & &  & \\
4-0 R(0) & 1001.83 & $<24.9$ & $<1.22$ \\
8-0 R(1) & 1002.46 & $70.3\pm12.5$ & $4.36\pm0.56$ \\
4-0 R(2) & 1051.51 & $24.2\pm8.6$ & $1.77\pm0.63$ \\
8-0 R(3) & 1006.42 & $<26.1$ & $<1.80$ \\
5-0 R(4) & 1044.54 & $<19.2$ & $<1.30$ \\
total\tablenotemark{b} &  & & $<11.64$ \\
\enddata
\tablenotetext{a}{Upper limits are 3$\sigma$.}
\tablenotetext{b}
{The upper limit on the total H$_2$ column density detected in the
{\em FUSE} spectra, derived by summing the $J=0-4$ levels.}
\end{deluxetable} 

The upper limits on the total H$_2$ column density are also shown in
Table 3. These were derived by simply summing the measured upper
limits (or detections) in the $J=0-4$ levels. In typical Milky Way ISM
conditions most ($>90$\%) of the H$_2$ is in the $J=0$ and $J=1$
states ({\it e.g.,} Snow et al. 2000, Friedman et al. 2000, Sembach et
al. 2001). If this is true in the starbursts as well, the upper $J$
levels should not contribute much to the total $H_2$ column
density. However, since the conditions in the ISM of the starbursts
may be different ({\it e.g.,} the potentially stronger UV radiation
field could populate the higher $J$ levels), summing all five levels
may be appropriate.

Table 4 lists an estimate of the upper limits on the mass of H$_2$
detectable by \fuse\ within the aperture. It is important to note that
this does not correspond to the total H$_2$ mass in the aperture,
because the majority of the molecular gas is too dense to be detected
by \fuse\ (see section 5.1). To determine this limit we first applied
the measured column density upper limit to the entire aperture, or in
other words we assumed the entire aperture is filled with H$_2$ at the
column density listed in Table 3. The angular area of the aperture was
converted to linear area at the assumed distance of the galaxy (see
Table 1), and multiplied by the column density limit. As we discuss in
detail below, these upper limits need to be treated carefully, given
the strong bias in these far-UV spectra against the dustiest
lines-of-sight.

The upper limits in Table 3 do not account for the possible presence
of unresolved, saturated absorption. Cold H$_2$ clouds with $T=100$~K
may have Doppler $b$-values $\la1$~\kms. A saturated absorption line
with a $b$-value of 1~\kms\ will have an equivalent width $EW\la10$
m\AA. Such a line could easily go undetected in the \fuse\ data
presented here, given the typical upper limits listed in
Table~3. Typical $b$-values measured for H$_2$ in the Milky Way and
Magellanic Clouds are much broader than this, ranging from 2.3 to 20
\kms\ (see Shull et al. 2000, Tumlinson et al. 2002). This broadening
is most likely due to bulk motions within individual clouds and the
velocity dispersion of the ensemble of clouds. If $b=3$~\kms,
$EW\ga17$~m\AA, which would result in a $2\sigma$ detection in most
cases. If unresolved, saturated absorption is present it would result
in the underestimation of the upper limits on the H$_2$ column density.

\subsection{Atomic Hydrogen}

Table 4 shows an estimate of (or lower limit on) the \ion{H}{1} column
density in the \fuse\ aperture. The \ion{H}{1} column was estimated in
different ways for each galaxy. \cite{heckman01} derived the
\ion{H}{1} abundance in NGC~1705 by fitting the Ly$\beta$ and
higher order \ion{H}{1} lines in the \fuse\ spectrum. For NGC~3310,
NGC~4124, and NGC~5253 we measured the equivalent width of the
\ion{Ar}{1} 1066.660~\AA\ line and converted this to a column density
assuming the line is optically thin. The \ion{Ar}{1} column density
was converted to \ion{H}{1} by assuming the relative Warm Neutral
Medium abundances given by \cite{sembach00}, and scaling down by the
O/H value given in Table 1. This line was used because dust depletion
is not important for \ion{Ar}{1} (e.g. Aloisi et al. 2003).  It is
also a relatively weak line, so while the assumption that it is
optically thin is likely invalid for most of these galaxies, it gives
a better lower limit to the column density than would a stronger line.
However, there can be a significant ionization correction for
\ion{Ar}{1}, and furthermore the O/H values in \ion{H}{2} regions (used
here) are usually higher than those in the neutral ISM (see {\it
e.g.,} Thuan et al. 2002; Aloisi et al. 2003). Thus, this approach
gives a conservative lower limit to N(\ion{H}{1}) for most of these
sight lines. The \ion{Ar}{1} line was not detected in the \fuse\
spectrum of M83, so we measured the
\ion{O}{1} 950.885~\AA\ line, and converted this to \ion{H}{1} using
the O/H value given in Table 1. This value is almost identical to the
upper limit on $N$(\ion{H}{1}) derived from the non-detection of the
\ion{Ar}{1} line, so we treat it as an estimate of the \ion{H}{1}
column density probed by the \fuse\ spectrum, not a lower limit. As we
will show in section 5.2, there are indications that the \ion{H}{1}
column densities are actually much larger than the conservative limits
derived above.

\begin{deluxetable*}{lcccccc}
\tabletypesize{\scriptsize}
\tablewidth{0pc}
\tablecaption{Diffuse H$_2$ Parameters of the Sample Galaxies} 
\tablehead{
\colhead{Galaxy} & \colhead{$N$(H~I)\tablenotemark{a}} & \colhead{$f$(H$_2$)\tablenotemark{b}} & \colhead{M(H$_2$)\tablenotemark{c}} & \colhead{M(H$_2$) from CO\tablenotemark{d}} & \colhead{I$_{1000}$\tablenotemark{e}} \\
\colhead{}       & \colhead{} & \colhead{}  & \colhead{(M$_{\odot}$)}  & \colhead{(M$_{\odot}$)}    & \colhead{(erg s$^{-1}$ cm$^{-2}$ \AA$^{-1}$ arcsec$^{-2}$) }     }
\startdata
NGC~1705       & $1.6\times10^{20}$  & $<4.9\times10^{-6}$ & $<3.5$   & \nodata             & $5.6\times10^{-16}$   \\
NGC~3310       & $>3.0\times10^{19}$  & $<2.9\times10^{-5}$ & $<30.8$ & $\sim10^8$          & $2.0\times10^{-16}$   \\
NGC~4214       & $>2.7\times10^{20}$  & $<6.1\times10^{-6}$ & $<2.4$  & $\sim8\times10^5$   & $1.3\times10^{-16}$   \\
M83 (NGC~5236) & $1.1\times10^{19}$  & $<1.6\times10^{-4}$ & $<6.1$  & $\sim10^7$          & $1.9\times10^{-16}$   \\
NGC~5253       & $>3.0\times10^{20}$  & $<7.8\times10^{-6}$ & $<0.4$  & $\sim10^5$          & $9.0\times10^{-16}$   \\
\enddata
\tablenotetext{a}
{H~I was determined by converting O~I or Ar~I absorption measurements
in the {\em FUSE} spectrum (assuming H~II region abundances), except
for NGC~1705 for which a measurement was available (Heckman et
al. 2001a)}

\tablenotetext{b}
{The H$_2$ fraction (see equation 1 in section 4) in the {\em FUSE}
aperture, using the upper limit on total $N$(H$_2$) from Table
3. These upper limits refer to the integrated light in the {\em FUSE}
aperture. There are undoubtedly regions within the aperture that have
higher molecular fractions.}

\tablenotetext{c}
{The upper limit on the total mass of H$_2$ in the {\em FUSE}
aperture, using the upper limit on total $N$(H$_2$) from Table 3. This
is only the mass of H$_2$ detectable with \fuse; most of these
galaxies contain much more mass in the form of dense clouds which are
invisible to \fuse\ (see text).}

\tablenotetext{d}
{An estimate of the H$_2$ mass within the {\em FUSE} aperture inferred
from CO measurements. See section 3 for references. }

\tablenotetext{e}
{The measured surface brightness (flux divided by aperture angular
area).}
\end{deluxetable*}

The third column in Table 4 lists the molecular hydrogen fractions
$f$(H$_2$) for the diffuse ISM in each galaxy, defined as:
\begin{equation}
f(\mathrm H_{2})=\frac{2N(\mathrm H_{2})}{[2N(\mathrm H_{2})+N(\mathrm H\mathrm I)]}.
\end{equation}
Since $N$(H$_2$) is an upper limit and $N$(\ion{H}{1}) is a lower
limit (for three of the galaxies), this expression gives an upper
limit to $f$(H$_2$).  The fact the we are almost certainly
underestimating $N$(\ion{H}{1}) implies that the true values of
$f$(H$_2$) could be as low as $10^{-7}$, and perhaps even lower if
$N$(H$_2$) is much below the upper limits in Table 3. However, it is
important to note that this upper limit applies to the integrated
light in the FUSE aperture, and that there certainly are regions
within the aperture, such as the molecular clouds seen in CO emission,
in which the molecular fraction is higher.

\section{Discussion}

\subsection{H$_2$ in the Dense ISM}

The $FUSE$ spectra show very little absorption from H$_2$ in the
starburst regions of these galaxies, with the upper limits on H$_2$
masses ranging from $\sim31$ to $<1$~M$_{\odot}$ (see Table
4). Similar results were found in $FUSE$ data for the metal-poor
starburst galaxy I~Zw~18 \citep{vidal00, aa03} and for the blue
compact dwarf galaxy Mrk~59 \citep{thuan02}. This is in stark contrast
with CO observations of starbursts, which often reveal large
concentrations of molecular gas in the starburst region. Published CO
studies of four of the five starbursts in our sample have found many
orders of magnitude more H$_2$ in these galaxies than the $FUSE$
measurements imply (see Table 4 and section 3).

The explanation for the difference between the emission and absorption
results is that the absorption spectra are not sensitive to the dense
clouds seen in CO emission. The combined effects of extinction and the
fact that the continuum source is extended biases the far-UV
observations toward the least dusty environments, because dust quickly
extinguishes the far-UV light in other regions (see {\it e.g.}  Bluhm
et al. 2003). Since the most efficient formation mechanism for H$_2$
is formation on dust grains \citep{hs71}, molecular gas is likely
concentrated in the dustiest regions. Dense molecular clouds have many
magnitudes of extinction in the far-UV, rendering them opaque to the
background continuum light. Since large amounts of CO are detected in
these galaxies, the lack of H$_2$ seen in absorption implies that most
of the molecular gas is in the form of dense clouds.  The continuum
that is visible in the \fuse\ spectra passes through the diffuse ISM
between the clouds. Since the \fuse\ spectra detect far-UV continuum
from these starbursts, the covering factor of the dense molecular
clouds in front of the UV continuum source must be $<1$.

Although it is tempting to consider the metallicity dependence of the
CO--H$_2$ conversion factor as the cause of the discrepancy between
the emission and absorption results, the magnitude of this effect is
not nearly large enough to account for the observed difference.
\cite{w95} found a factor of 4.6 increase in the conversion factor for
a factor of 10 decrease in metallicity. The metallicity range of the
sample is within a factor of 10 from solar abundances, so the
discrepancy between the \fuse\ and CO measurements is therefore much
too large to be caused by metallicity. Furthermore, metal-poor
galaxies such as NGC~5253 should have {\it more} H$_2$ than the CO
measurements imply, which is the opposite of what we observe.

\subsection{H$_2$ in the Diffuse ISM}

Sight lines with diffuse H$_2$ column densities ranging from
$10^{14}-10^{21}$~cm$^{-2}$ are ubiquitous in the Milky Way and
Magellanic Clouds \citep{s77,shull00, t02}. Indeed, it is difficult to
find a sight line that extends more than $\sim1$~kpc that does not
show absorption from H$_2$. Absorption from H$_2$ has also been
detected in \fuse\ spectra of \ion{H}{2} regions in M33
\citep{b03}. It is thus surprising to see so little H$_2$ absorption
in the starburst spectra. In this section we discuss how observational
biases may explain the results, but also give evidence arguing that
the lack of H$_2$ absorption in the \fuse\ spectra points to a real
difference in the diffuse ISM in starbursts.

\subsubsection{Observational Selection Effects}

It is possible that the biases discussed in the previous section also
prevent us from detecting H$_2$ in the diffuse ISM. Simply put, if the
spatially extended continuum source (the starburst) is thought of as
many individual sight lines within the aperture, then those sight
lines with the least extinction (and therefore the lowest H$_2$ column
density) will contribute the most to the composite continuum. {\em
Copernicus} observations of sight lines to hot stars in the Milky Way
show a pronounced increase in the molecular hydrogen column density
occurring at $E(B-V)=0.08$, corresponding to
$N$(H$_2$)$=10^{19}$~cm$^{-2}$ (Savage et al. 1977; also see
Figure~\ref{savage}). \cite{bohlin78} showed that the amount of total
hydrogen (\ion{H}{1}$+$H$_2$) that corresponds to this amount of
reddening ($\sim5\times10^{20}$~cm$^{-2}$) is enough to shield H$_2$
from destruction by UV photons, explaining the marked increase in
column density. If the continuum at 1000~\AA\ is dominated by sight
lines with $E(B-V)<0.08$, it is possible that we would not see any
H$_2$ absorption since these sight lines are the least likely to
contain H$_2$. Indeed, if the lower limits on the \ion{H}{1} column
densities in Table 4 are close to the true values, they would suggest
that the sight lines probed by the \fuse\ spectra lack the shielding
necessary for the survival of H$_2$. The \fuse\ data by themselves
seem consistent with the {\em Copernicus} results for the Galaxy.

\subsubsection{Arguments for a Deficiency of Diffuse H$_2$}

\begin{figure}
\epsscale{1.19}
\plotone{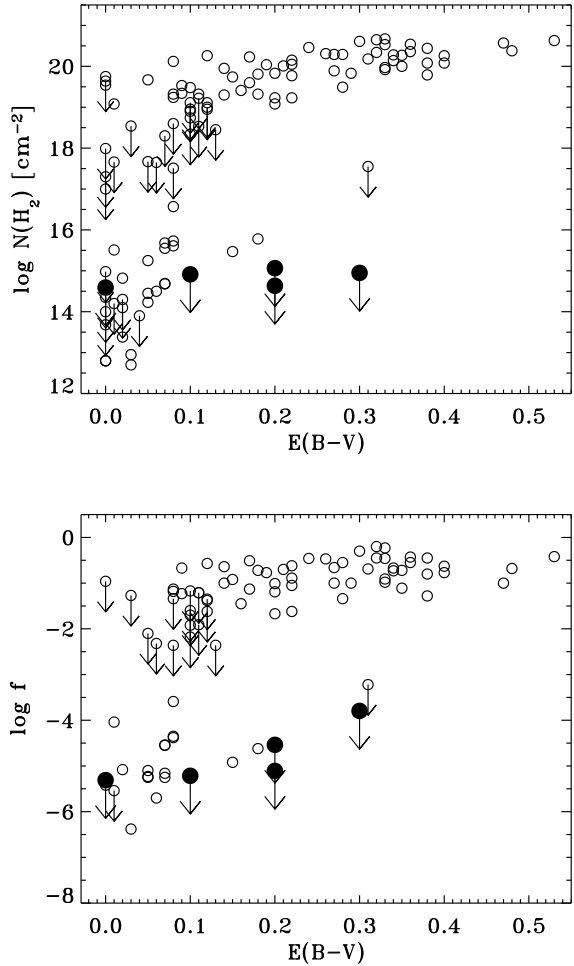}
\caption{
The H$_2$ column density (top panel) and molecular fraction $f$
(bottom panel) versus reddening. The solid symbols are the {\em FUSE}
starburst upper limits, and the open symbols are {\em Copernicus}
measurements and limits for Galactic sight lines (from Savage et
al. 1977). The H$_2$ column densities and molecular fractions in the
starbursts are lower than the Galactic sight lines for similar
reddenings (except for NGC~1705). Note that the molecular fractions
for the starbursts are probably much lower than the derived upper
limits due to the probable underestimation of the HI column
density. The three highly-reddened Galactic sight lines marked with
double circles were noted by Savage et al. (1977) as possibly having
strong radiation fields.}
\label{savage}
\end{figure}

Because the lower limits on $N$(\ion{H}{1}) given in Table 4 are based
on lines that are very likely saturated, they may severely
underestimate the true column density. Given this limitation, it may
be more useful to compare the average {\it reddening} in the starburst
sight lines with the Milky Way. Spectra of these starbursts taken with
the {\em International Ultraviolet Explorer (IUE)} and {\em Hopkins
Ultraviolet Telescope (HUT)} show that the reddening within these
galaxies is substantial (Meurer, Heckman, \& Calzetti 1999; Leitherer
et al. 2002).  As listed in Table 1, $E(B-V)$ is $\sim$0 in NGC~1705,
but ranges from 0.2 to 0.3 in the other four galaxies. Scaling to the
observed metallicities in the starbursts assuming a Milky Way ratio of
dust/metals in the ISM, the implied \ion{H}{1} column densities range
from one to several $\times10^{21}$~cm$^{-2}$ in these four cases.

\begin{figure*}
\epsscale{1.02}
\plotone{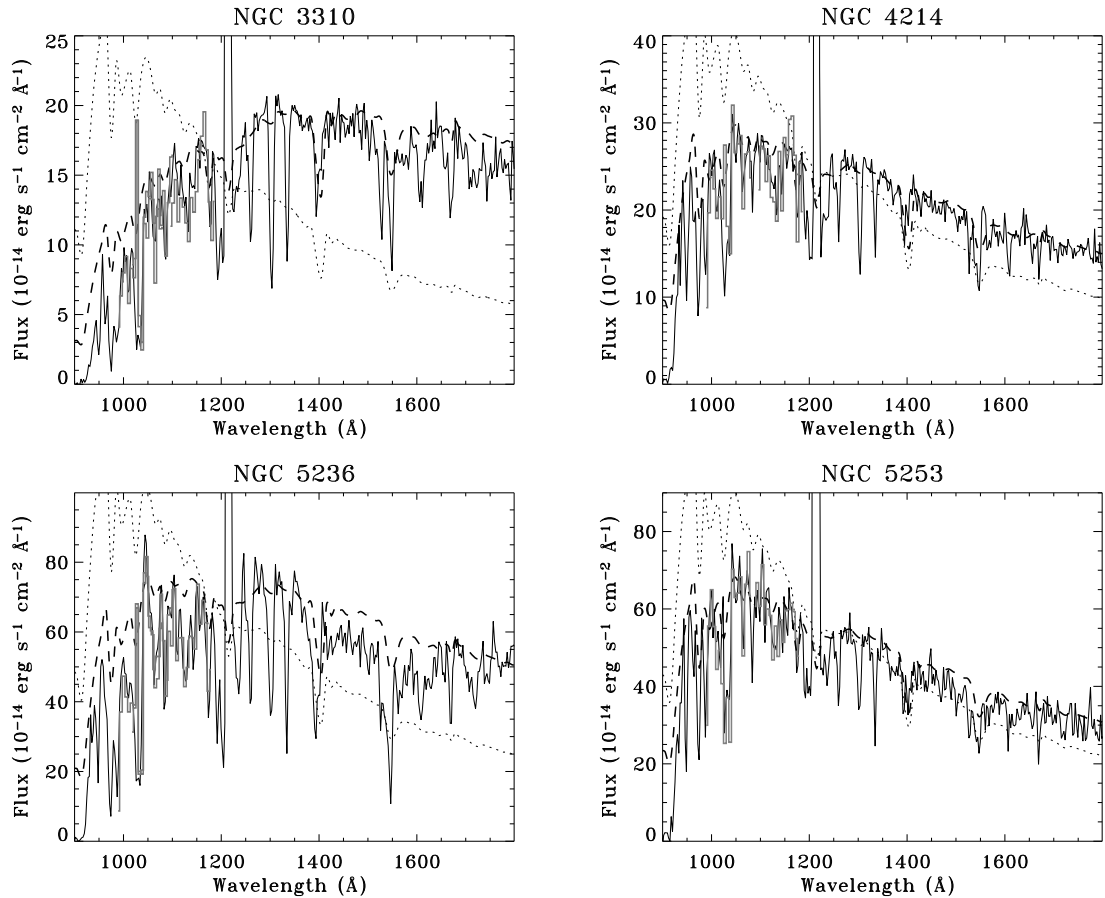}
\caption{
The {\em Hopkins Ultraviolet Telescope (HUT)} spectra of the four most
reddened galaxies in our sample (NGC~1705 is excluded). The {\em FUSE}
spectra are also shown as gray thick lines. The dotted line in each
panel is the SED of a 100 Myr continuous star formation model, from
Starburst99 (Leitherer et al. 1999). The dashed line shows the same
SED reddened using the starburst extinction law as given in Leitherer
et al. (2002). The reddening values applied were chosen to give the
best fit (by eye) to the observed {\em HUT} spectra, and the resulting
$E(B-V)$ values are all within $\pm0.1$ of those listed in Table
1. The models match the data reasonably well, even near 1000~\AA,
where the H$_2$ measurements were made. In some of the panels the
attenuation at 1000~\AA\ is underestimated, implying that there is
more dust/gas along the sight line than the reddening values in Table
1 would indicate. This figure illustrates that the observed spectra
are consistent with $E(B-V)$ values of at least $0.1-0.3$ at 1000~\AA,
and are not consistent with zero reddening.  }
\label{hut}
\end{figure*}

While the \fuse\ spectra are certainly biased toward the sight lines
with the lowest amount of reddening, it still appears that the
majority of sight lines that contribute to the UV flux are reddened by
$E(B-V)=0.2-0.3$ (except for NGC~1705). Now we can directly compare
the starburst sight lines with those in the Milky Way. \citet{s77}
compared $N$(H$_2$) and $E(B-V)$ for Milky Way sight lines. This
comparison is reproduced in Figure~\ref{savage} with the starburst
points included. The H$_2$ column densities and molecular fractions in
the starbursts are lower than the Galactic sight lines for similar
amounts of reddening (except for NGC~1705). While there is uncertainty
in the amount of \ion{H}{1} along the sight lines, the reddening
values, measured in the UV, are well within the range of H$_2$ bearing
sight lines in the Milky Way. Note that the molecular fractions ($f$)
for the starbursts are probably much lower than the derived upper
limits if the measured $E(B-V)$ values are correct and a Milky Way
gas/dust ratio is assumed.  Thus, the difference between the
starbursts and the Milky Way sight lines in Figure~\ref{savage} may be
even more pronounced. Figure~\ref{savage} suggests that there is a
deficiency of H$_2$ outside of dense clouds in starbursts compared to
the Milky Way.

This interpretation is consistent with the lower limits on
$N$(\ion{H}{1}) and the measured $E(B-V)$. Indeed, if there is a Milky
Way $H_2$ abundance in the starbursts, there must be a difference
between the $E(B-V)$ values in the {\em IUE} and \fuse\ bands. It is
possible to envision scenarios where the actual extinction at
1000~\AA\ is smaller than that measured in the $1200-1800$~\AA\
spectral range of {\em IUE}, since the rise in the extinction curve
causes the most reddened sight lines to contribute proportionally less
to the total flux at shorter wavelengths. However, comparing the
entire {\em HUT} spectrum with reddened model starburst SEDs shows a
very good match even in the $1000-1200$~\AA\ {\em FUSE} range (see
Figure~\ref{hut}), which would not be the case if the spectrum at
1000~\AA\ were preferentially unreddened compared to the longer
wavelength data. The generic SEDs in Figure \ref{hut} match the data
very well even though we did not attempt to match the models to the
metallicity or stellar content of the galaxies. This strongly suggests
that there is still substantial reddening at 1000~\AA, with
$E(B-V)=0.1-0.3$.

Another factor to take into account is the possibility that there is a
significant amount of ionized gas along the sight
lines. \cite{heckman01} showed that this is true for NGC~1705. Figure
6 shows that the \ion{Si}{2} absorption in M83 is much weaker than in
the other galaxies (except NGC~1705), implying a lower column of
neutral gas.  The $N$(\ion{H}{1}) measurement is only
$\sim10^{19}$~cm$^{-2}$, two orders of magnitude smaller than the
\ion{H}{1} column implied by the reddening.  The discrepancy between
the reddening and the neutral gas column may indicate that most of the
hydrogen in the diffuse ISM along the sight line is ionized. Studies of
sight lines in the Galaxy have shown that the dust content of the
ionized ISM is similar to that in the neutral gas \citep{howk99}.

If extinction limited the \fuse\ spectra to only a few nearly
unreddened sight lines, the continuum would contain the light of a
small number of stars, and they would perhaps only probe the outer
edge of the galaxy. This is ruled out by the spectra themselves. The
stellar features seen in the far-UV continuum in all cases is
dominated by light from a very young population of O and B stars
(e.g. Leitherer et al. 2002; Robert et al.  2003; Leitherer et al. in
preparation) and the UV fluxes imply the presence of $\sim1000$ such
sources (or more if they are reddened). Furthermore, the kinematic
signature of the starburst outflow is also clearly evident in the
\ion{O}{6} lines \citep{heckman01,hsmlcm01}, which indicates that the
continuum source is the starburst, rather than just a few OB stars in
the outer halos of the galaxies.

To summarize, the upper limits on $N$(H$_2$) derived from the \fuse\
starburst spectra are consistent with the \citet{s77} Milky Way
relationship between $N$(H$_2$) and $N$(\ion{H}{1}) in the diffuse
ISM. They cannot rule out, however, a deficiency of $N$(H$_2$)
compared to Milky Way sight lines with similar amounts of reddening in
four of the galaxies. Indeed, when the {\it IUE} and \hut\ measured
reddening is included in the analysis, the data point to such a
deficiency. However, with the differences in the \iue, \hut, and
\fuse\ apertures, and the possibility of differences in the reddening
measured at \fuse\ versus \iue\ wavelengths, the present data cannot
definitively rule out a Milky Way H$_2$ abundance in the diffuse ISM
of these starbursts.

\subsection{Formation and Destruction of H$_2$}

If the suggested lack of H$_2$ in the diffuse ISM of starbursts is
real, there are implications for H$_2$ formation and destruction
mechanisms. Low column densities of H$_2$ in the diffuse ISM could be
caused by either inhibited formation or rapid destruction of
H$_2$. \cite{t02} and Browning, Tumlinson, \& Shull (2003) calculated
the effects of a stronger radiation field and a lower H$_2$ formation
rate on $f$(H$_2$). To reproduce such low $f$(H$_2$) values at the
observed \ion{H}{1} column densities requires a radiation field at
least 50 times stronger than the Galactic mean UV radiation field, or
a formation rate less than 1/10th that of the Milky Way, or a
combination of both effects (see Figure 4 in Browning et al. 2003).

Inhibited formation is an attractive scenario in metal-poor galaxies
because of the deficiency of dust grains on which to form
molecules. While the most efficient mechanism for H$_2$ formation is
on the surfaces of dust grains \citep{hs71}, other mechanisms have
been proposed to occur in the absence of dust \citep{jp97}. However,
the fact that the \fuse\ spectra show no H$_2$ absorption in galaxies
where H$_2$ is known to exist indicates that the H$_2$ is associated
with dust. This suggests that formation on the surface of grains is
the dominant mechanism in these galaxies.

\cite{vidal00} and \cite{aa03} used \fuse\ to search for
H$_2$ absorption in I~Zw~18, and \cite{thuan02} did the same for
Mrk~59. These are both metal-poor galaxies undergoing star
formation. They set low upper limits on the abundance of H$_2$ in the
diffuse ISM, similar to those we find for our sample of
starbursts. Because I~Zw~18 and Mrk~59 have not been detected in CO it
is not known whether these galaxies contain any H$_2$, even in the
form of dense clouds. Thus it was not clear to what extent the low
metallicity inhibited the formation of molecular gas. Our sample spans
a wide range of metallicity, and the CO detections in most of the
galaxies in our sample establish that large amounts of molecular gas
have formed. This suggests that inhibited formation is not the only
factor leading to the low H$_2$ content in the diffuse
ISM. Furthermore, M83 and NGC~3310 have high metal abundances and are
very bright far-infrared sources \citep{calzetti95}, indicating the
presence of dust. Yet we still detect very little H$_2$ outside of
dense molecular clouds, ruling out metallicity as the culprit for
those galaxies at least.

The remaining scenario then is rapid destruction. In this scenario
molecules of H$_2$ that evaporate from the surfaces of the dense
clouds are photodissociated by UV radiation \citep{sw67}. The intense
UV radiation fields of starburst galaxies may create an environment
where the destruction of H$_2$ would be enhanced. \cite{bts03} modeled
the formation and destruction of H$_2$ in the Milky Way and Magellanic
Clouds. They assume
$I=1\times10^8$~photons~cm$^{-2}$~s$^{-1}$~Hz$^{-1}$ for the Galactic
radiation field, which corresponds to a surface brightness
$1.1\times10^{-18}$~ergs~cm$^{-2}$~s$^{-1}$~\AA$^{-1}$~arcsec$^{-2}$
at 1000~\AA. Table 4 lists the surface brightness at 1000~\AA\ for
each galaxy, measured from the \fuse\ spectra by dividing the flux by
the angular area of the aperture.  All are more than 100 times
stronger than the Galactic radiation field, neglecting any possible
effects of extinction. This is an average across the aperture, and
since the aperture is not uniformly filled with UV continuum the
radiation field is undoubtedly stronger at some locations. According
to the \cite{bts03} models this is more than enough to produce the low
$f$(H$_2$) values seen in these galaxies, even if there is no
reduction in the formation rate.  In addition to the effects of UV
radiation, the multiple supernovae that have occurred in these
galaxies produce fast shocks in the ISM, which can also enhance the
dissociation of molecular gas. In such a harsh environment it would
not be surprising to find that there is too little H$_2$ in the
diffuse ISM to detect. It is interesting to note that the three
highly-reddened Galactic sight lines marked as double circles in
Figure~\ref{savage} ($\theta^1$ Ori C, 29 CMa, 30 CMa) were noted by
Savage et al. (1977) for their abnormally low H$_2$ column densities,
possibly due to an unusually strong dissociating radiation
field. These sight lines lie the closest to the starbursts in the
figure.

\subsection{Dust in the Diffuse ISM}

We have given arguments that there is very little H$_2$ in the diffuse
ISM of these starbursts. We have also shown that the ultraviolet
radiation field in these galaxies is strong enough to destroy H$_2$ in
the diffuse ISM. However, as previously discussed, UV spectroscopy has
shown that there is substantial reddening of the UV continuum,
indicating the presence of dust in the diffuse ISM of these galaxies
(except perhaps for NGC~1705, which is close to unreddened).  Thus,
the molecular gas and dust may have different distributions in these
galaxies: H$_2$ is confined to dense clumps that are highly optically
thick to the far-UV continuum, and is absent in the diffuse
ISM. However, dust grains exist in both dense and diffuse
environments. If the starbursts are deficient in diffuse H$_2$, the
mechanism responsible for dissociating H$_2$ (UV radiation field,
shocks, or both) apparently is not strong enough to destroy all of the
dust.

These arguments also support the idea that the covering factor of the
dense clumps in front of the UV continuum source is
small. \cite{meurer99} found an empirical relation between spectral
slope $\beta$ and the ratio of far-infrared (FIR) to UV fluxes, such
that galaxies with strong FIR flux relative to UV flux have shallower
(redder) UV spectra. The FIR excess can be attributed to dust
extinction in the UV, so their conclusion was that starbursts with
high extinction are also reddened. This relationship can only exist if
the covering factor of dense clumps is small, because these clumps
produce FIR flux but do not redden the UV spectrum since they are
opaque to UV continuum. The picture based on theoretical modeling of
the \cite{meurer99} result by \cite{wg00}, \cite{cf00}, and others
(see Calzetti 2001 for a recent review) is that the strong correlation
between the UV color and the ratio of far-IR to UV flux can only be
understood in the context of an inhomogeneous dusty medium lying in
the foreground between the UV sources and observer. Our results imply
that this medium (which is translucent to far-UV photons) contains
substantial amounts of dust but not molecular gas.

NGC~1705 is an interesting case: its UV spectral slope is close to
that of an unreddened young stellar population \citep{meurer99}, and
unlike the other galaxies in our sample it is not a strong
far-infrared source \citep{calzetti95}. These facts suggest that it
contains less dust than the other starbursts in the sample, similar to
I~Zw~18 and Mrk~59. However, we cannot yet say that these galaxies are
devoid of molecular gas. \cite{cannon02} used narrow-band imaging to
map the dust content of I~Zw~18. The dust mass is consistent with the
low-metallicity and low FIR emission, and the morphology is very
clumpy. If the dense H$_2$ is traced by dust, this clumpiness would
explain the non-detection with \fuse. It seems likely that there is
molecular gas in I~Zw~18, even though it has not been detected in CO
emission (perhaps due to low metallicity) or H$_2$ absorption (due to
the clumpy morphology). The same may hold true for NGC~1705 and
Mrk~59.

\subsection{Further Implications}

\cite{r02} carried out a survey of pure rotational emission from H$_2$
in starbursts (including M83 and NGC~5253) using the {\em Infrared
Space Observatory}. This emission traces warmer H$_2$ ($T\sim150$~K)
than that typically seen in CO emission ($T\la100$~K). At least some
of the warm gas is thought to be in the photodissociation regions at
the surfaces of cold clouds, although there is also evidence for an
extended warm component \citep{v99}. The survey found that the warm
H$_2$ makes up $1-10$\% of the total H$_2$ content derived from CO
observations. This is still much higher than the limits on H$_2$ in
the diffuse ISM set with the \fuse\ spectra. Either the column density
of an extended warm component is high enough that it is opaque to
far-UV light, or the bulk of the warm gas is associated with the dense
clouds of cold H$_2$, suggesting perhaps that is in photodissociation
regions.

The low limits on $f$(H$_2$) found for these starbursts are
reminiscent of the values or upper limits measured in damped
Ly$\alpha$ systems (DLAs) in QSO spectra (see {\it e. g.,} Levshakov et
al. 2002 and references therein). The difference is that in the
starbursts we know there is much more H$_2$ than is apparent in the
\fuse\ spectra, indicating that the spectra are not sensitive to regions
in the aperture containing dense H$_2$. A similar effect may hinder
the detection of DLAs with H$_2$ if the dust associated with the H$_2$
in such DLAs makes the QSOs behind them invisible ({\it e.g.,} Fall \&
Pei 1993).  The known QSOs would then preferentially probe DLAs with
little extinction, and thus little molecular gas, or systems where the
line of sight to the QSO happens to pass through a hole in a clumpy or
filamentary H$_2$/dust distribution \citep{hfwr03}. \cite{ellison01}
found that the underestimate of the population of DLAs due to dust is
at most a factor of 2. However, the DLAs that are missed are those
with the most H$_2$ in the line of sight. This would result in an
underestimate of the typical H$_2$ content of DLAs.

\section{Conclusions}

We have searched for absorption from H$_2$ associated with the diffuse
ISM in \fuse\ spectra of five starburst galaxies. We have tentatively
detected H$_2$ in M83 and NGC~5253, and set upper limits for NGC~1705,
NGC~4214, and NGC~3310. In general there is much less H$_2$ seen in
the far-UV than implied by previous mm-wave CO measurements, as
expected if most of the H$_2$ is in dense clumps which \fuse\ cannot
detect because they are opaque to far-UV light. The upper limits on
$N$(H$_2$) and the lower limits on $N$(\ion{H}{1}) in the starbursts
are consistent with the \citet{s77} values for the diffuse ISM in the
Milky Way. However, the substantial reddening measured in {\it IUE}
spectra of four of the starbursts suggests that the amount of H$_2$ in
the diffuse ISM is much lower than for sight lines in the Milky Way or
Magellanic Clouds with similar amounts of reddening. If the suggested
deficiency of diffuse H$_2$ is real, it is likely due to
photodissociation of H$_2$ molecules by UV radiation from massive
stars or shocks from multiple supernovae, and illustrates how the
harsh environment in starbursts affects the ISM. However, the
mechanism that dissociates the molecular gas apparently does not
destroy the dust which reddens the UV spectra. The \fuse\ observations
show that the absence of H$_2$ absorption in the far-UV does not
necessarily indicate a lack of molecular gas in an actively
star-forming galaxy.

\acknowledgments

We thank the anonymous referee for useful, constructive comments.
This work benefited from discussions with Lynn Hacker, B-G Andersson,
Bill Blair, Paul Feldman, and Stephen McCandliss.  This research has
made use of the NASA/IPAC Extragalactic Database (NED) which is
operated by the Jet Propulsion Laboratory, California Institute of
Technology, under contract with the National Aeronautics and Space
Administration. This project was supported by NASA grant NAG5-9012.

\end{document}